\documentclass{ws-ijmpd}
\usepackage[super,compress]{cite}
\usepackage{amsmath}
\usepackage{ulem}
\usepackage{xcolor}

\def\b{\begin{equation}}
\def\e{\end{equation}}
\def\be{\begin{eqnarray}}
\def\ee{\end{eqnarray}}

\begin{document}

\markboth{Benone, Leite, Crispino, Dolan}
{On-Axis scalar absorption cross section of Kerr-Newman black holes}

%
\catchline{}{}{}{}{}
%

\title{On-Axis scalar absorption cross section of Kerr-Newman black holes: Geodesic analysis, 
sinc and low-frequency approximations}

\author{CAROLINA L. BENONE, LUIZ C. S. LEITE AND LU\'IS C. B. CRISPINO}
\address{Faculdade de F\'{\i}sica, Universidade Federal do Par\'a\\ 66075-110, Bel\'em, Par\'a, Brazil\\
lben.carol@gmail.com, luizcsleite@ufpa.br, crispino@ufpa.br}

\author{SAM R. DOLAN}
\address{Consortium for Fundamental Physics, School of Mathematics and Statistics,
University of Sheffield, Hicks Building, Hounsfield Road, Sheffield S3 7RH, United Kingdom\\s.dolan@sheffield.ac.uk}

\maketitle


\begin{abstract}
We  investigate null geodesics impinging parallel to the rotation axis of a Kerr-Newman black hole, and show that the absorption cross section for a massless scalar field in the eikonal limit can be described in terms of the photon orbit parameters. We compare our sinc and low-frequency approximations with numerical results, showing that they are in excellent agreement.
\end{abstract}

\keywords{Kerr-Newman black holes; null geodesics; scalar absorption.}

\ccode{PACS numbers: 04.50.-h, 04.50.Kd, 04.20.Jb}


\section{Introduction}	

Geodesics are a major issue of study in general relativity, since they give us the path followed by free particles in a spacetime. They help us to understand the structure of the spacetime, revealing new features. For geometries with rotation, the study of geodesics shows the presence of an ergoregion, {\it i. e.} a region where the particle has to rotate in the same sense as the central object. One can also study the formation of black holes by following the horizon generators, which follow null geodesics.

The curvature of the spacetime can act as a lens, making the path of light deviate from a straight line. This effect is related to the first experimental test of general relativity, and to the measurements of the deflection of light by the sun during the eclipse of 29th May 1919 \cite{Dyson:1920cwa}. This gravitational lensing makes it possible to see multiple images from a single source (e.g Einstein cross). For some cases the light of the source can be even lensed in a ring, in which case we have an Einstein ring\cite{Hewitt:1988}. 

By studying null geodesics one can also determine the shadow of a black hole. This shadow is expected to be seen in the near future with the Event Horizon Telescope \cite{Falcke:1999pj,Lu:2014zja}. Depending on the form of the shadow, we can differentiate between various kinds of black holes, which can also be used to test the no-hair theorem \cite{Cunha:2015yba}.

The importance of studying geodesics extrapolates the limit of classical physics, being important also to understand the behavior of quantum fields around black holes. For instance, one can show that the total absorption cross section of quantum fields by black holes presents regular oscillations around the capture cross section, which is given by $\sigma_{\text{geo}}=\pi b_c^2$, where $b_c$ is the critical impact parameter. Sanchez first noticed this feature when she studied the absorption spectrum of a massless scalar field around a Schwarzschild black hole \cite{Sanchez:1977si}. From this result she proposed a model involving a $\textrm{sinc}(x) = \sin(x)/x$ function, matching the parameters in order to fit the numerical curves. 
%

In Ref.~\refcite{Decanini:2011xi}, Sanchez's result was generalized for static black holes with any dimension. 
The authors used Regge pole techniques to show that the oscillatory pattern of the absorption cross section can be matched through a $\textrm{sinc}(x)$ function, where the parameters involved are related to the light ring. 
Recently, Macedo and collaborators showed that this result can be extended for the case of absorption by a Kerr black hole \cite{KerrAbsSca}. 
The Regge pole techniques used  in Ref.~\refcite{Decanini:2011xi} were also applied to charged black holes, whose absorption properties have been extensively investigated in the literature~\cite{CO:2008-CHO:2009, CHM:2010, OCH:2011, Benoneetal:2014-2017}.

In a recent paper the absorption cross section for a massless scalar field impinging upon a charged rotating black hole for different incidence angles has been investigated~\cite{Leite:2017hkm}. 
It has been shown that, for the on-axis case, the total absorption cross section oscillates around the classical limit, but the sinc approximation for this case was still lacking.
 
Besides the sinc approximation, in the present paper we investigate on-axis null geodesics in a Kerr-Newman black hole to find the high-frequency limit of the scalar absorption cross section. 
We organize the rest of this paper as follows: 
In section \ref{sec_noe} we consider null geodesics, finding analytical expressions for the on-axis case. In section \ref{sec_sa} we use the Regge pole technique to find an expression for the high-frequency limit of the absorption cross section. In section \ref{sec_nr} we present the equations for a massless scalar field impinging upon a Kerr-Newman black hole, describing the numerical procedure we used to find the solution, which we compare with the results for the sinc approximation. In section \ref{sec_lf} we reobtain the result for the low frequency limit, showing that it goes to the area of the black hole event horizon. In section \ref{sec_fr} we present our final remarks.

\section{Null orbit equations}
\label{sec_noe}
The Kerr-Newman black hole can be described by the  following line element
\be
d s^2&=& \left(1-\frac{2Mr-Q^2}{\rho^2}\right)d t^2-\frac{\rho^2}{\Delta}d r^2-\rho^2 d\theta^2 \nonumber\\
&+&\frac{4Mar\sin^2\theta-2aQ^2\sin^2\theta}{\rho^2}d t d\phi-\frac{\xi\sin^2\theta}{\rho^2}d\phi^2,\label{eq:linelement}
\ee
where ~$\rho^2\equiv r^2+a^2\cos^2\theta$, $\Delta\equiv r^2-2Mr+a^2+Q^2$, and $\xi\equiv (r^2+a^2)^2-\Delta a^2\sin^2\theta$. This line element corresponds to a black hole with mass $M$, charge $Q$ and angular momentum $J=aM$, provided that $a^2+Q^2 \leq M^2$. This black hole has an event horizon located at $r_+=M+\sqrt{M^2-(a^2+Q^2)}$ and a Cauchy horizon at $r_-=M-\sqrt{M^2-(a^2+Q^2)}$.

For this case, the equations of motion for a light ray are given by \cite{Carter:1968rr}
\be
&\rho^2& \dot{t} = \frac{(r^2+a^2)^2 - a \mathcal{L}_z (r^2+a^2)}{\Delta} - \sin^2{\theta}\left(a^2-\frac{\mathcal{L}_z a}{\sin^2{\theta}}\right),\label{eqt}\\
&\rho^4& \dot{r}^2 = [(r^2+a^2)-a \mathcal{L}_z]^2 - \Delta[(\mathcal{L}_z-a)^2+\mathcal{K}]\equiv R(r),\label{eqr}\\
&\rho^2& \dot{\phi} = \frac{a(r^2+a^2)-a^2\mathcal{L}_z}{\Delta} - \frac{a\sin^2{\theta}-\mathcal{L}_z}{\sin^2{\theta}},\label{eqphi}\\
&\rho^4& \dot{\theta}^2 = [\mathcal{K}+(\mathcal{L}_z-a)^2] -\frac{(a\sin^2{\theta}-\mathcal{L}_z)^2}{\sin^2{\theta}},\label{eqthe}
\ee
where $\mathcal{L}_z=L_z/E$ is the azimuthal angular momentum per unit energy and $\mathcal{K}=K_C/E^2$ is the Carter's constant ($K_C$) per unit energy ($E$) squared.

\begin{figure}
\centering
\includegraphics[width=1\columnwidth]{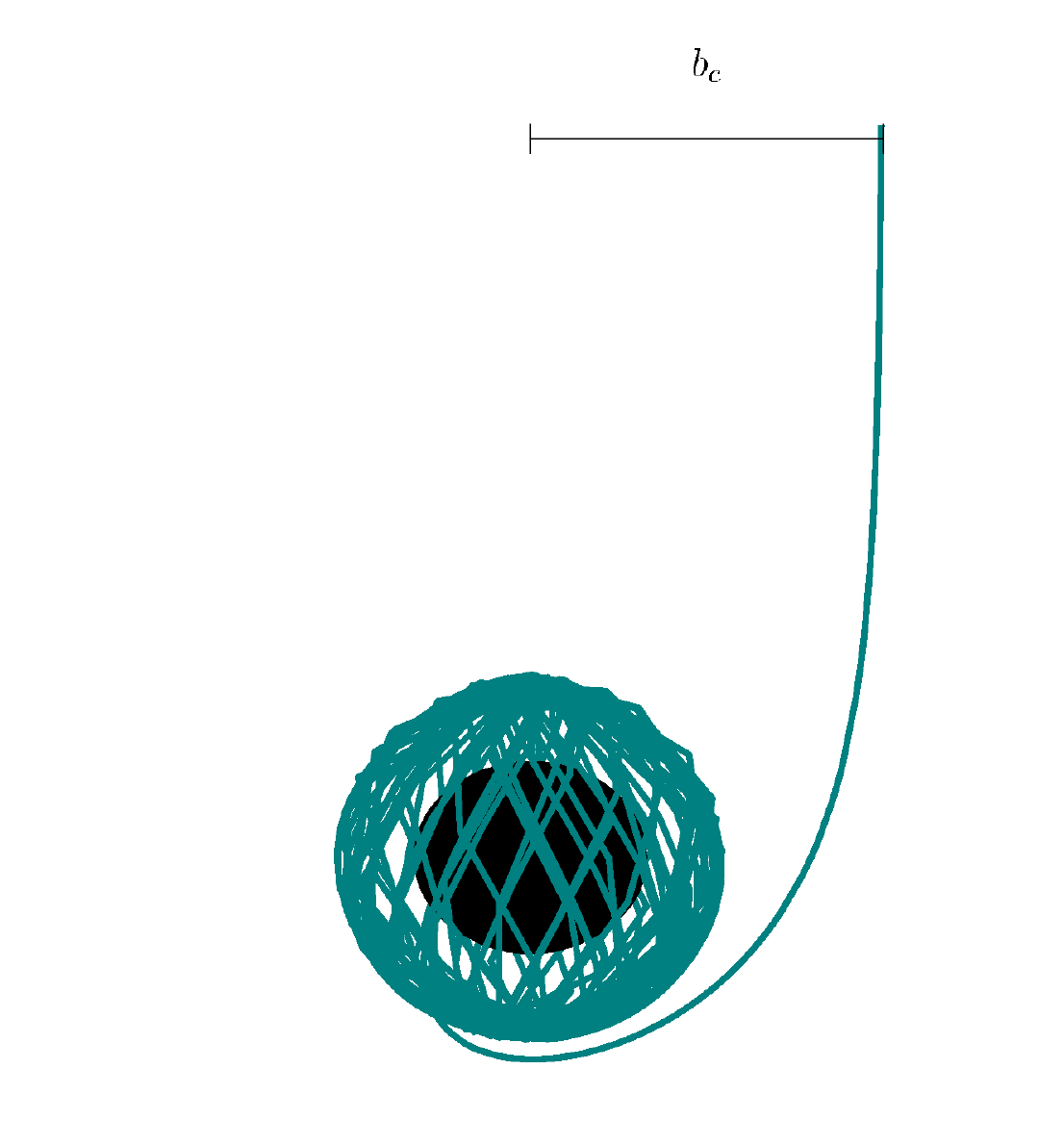}
\caption{Critical null geodesic with critical impact parameter ($b_c$) 
impinging parallel to the rotational axis of a
Kerr-Newman black hole with $a=0.9M$ and $Q=0.3M$.}
\label{npg}
\end{figure}

We will focus on the case of incidence along the black hole rotation axis, which implies $L_z=0$. For this case, the impact parameter $b$ is given by \cite{Dolan:2010wr}
\b
b=\sqrt{\mathcal{K}+a^2}.
\e
To find the circular null orbit we must solve $R(r_c)=0$ and $R'(r_c)=0$, in order to find the critical impact parameter $b_c$ and the critical radius $r_c$. For the critical impact parameter, we find
\b
b_c^2=\frac{2r_c(r_c^2+a^2)}{r_c-M}.
\e
As for the critical radius, we obtain
\b
r_c=M+2\sqrt{M^2-(a^2+2Q^2)/3}\cos\left[\frac{1}{3}\arccos\left[\frac{M^3-(a^2+Q^2)M}{(M^2-(a^2+2Q^2)/3)^{3/2}}\right]\right].
\e
In Fig. \ref{npg} we display a geodesic with critical impact parameter for a black hole with $a=0.9M$ and $Q=0.3M$. The total angular momentum is not conserved for the rotating case, which implies that the geodesic does not stay in the same plane. 

The time for a light ray to leave and return to the same pole, that is to undertake a latitudinal angle of $2\pi$, is called latitudinal period and can be found through
\b
T_0=2\int_{-1}^1 \frac{\dot{t}}{\dot{z}}dz,
\e
where $z=\cos{\theta}$. From Eqs. (\ref{eqt}) and (\ref{eqthe}) we find
\b
\dot{t}=\rho^{-2}\left[\frac{(r^2+a^2)^2}{\Delta} - (1-z^2)a^2\right]\biggr\rvert_{r=r_{c}},
\e
and
\b
\dot{z}= \left[\rho^{-2}\sqrt{(\mathcal{K}+a^2 z^2)(1-z^2)}\right]\biggr\rvert_{r=r_{c}},
\e
which gives us
\b
T_0 = \frac{4}{\mathcal{K}}\left\{\mathcal{K} \mbox{E}\left(-\frac{a^2}{\mathcal{K}}\right)+\left[\frac{(r^2+a^2)^2}{\Delta}-(a^2+\mathcal{K})\right]\mbox{K}\left(-\frac{a^2}{\mathcal{K}}\right) \right\},
\e
where $\mbox{K}(k)$ and $\mbox{E}(k)$ are the complete elliptic integrals of the first and second kind, respectively.

The Lyapunov coefficient is given by 
\b
\Lambda=\frac{1}{\dot{t}}\sqrt{\frac{1}{2}\frac{d^2V}{dr^2}},
\e
where $V=\dot{r}^2$. We can take an orbital average of this coefficient~\cite{Dolan:2010wr}, given by 
\b
\Lambda_0= \frac{2}{T_0}\int_{-1}^1 \bar{\lambda}\frac{\dot{t}}{\dot{z}}dz,
\e
such that we obtain
\b
\Lambda_0 = \frac{4\sqrt{6r^2+2a^2-b^2}}{T_0\sqrt{\mathcal{K}}}\mbox{K}\left(-\frac{a^2}{\mathcal{K}}\right).
 \e

\section{Sinc Approximation}
\label{sec_sa}

The total massless scalar absorption cross section for a charged rotating black hole 
is given by \cite{Leite:2017hkm}
\b
\sigma=\sum_{l=0}^\infty\sum_{m=-l}^l \frac{4\pi^2}{\omega^2}|S_{\omega l m}|^2 \Gamma_{\omega l m},
\label{abs}
\e
where $S_{\omega l m}$ are the oblate spheroidal harmonics~\cite{spheroidal} and $\Gamma_{\omega l m}$ are the greybody factors. By using the complex angular momentum technique we are able to find an analytic formula for the absorption cross section in the high-frequency limit in terms of the properties of the photon orbit. We make the transformation $l=\lambda-1/2$ in Eq. (\ref{abs}), where $\lambda$ is a complex number. Manipulating this equation, following closely Ref. \refcite{Decanini:2011xi}, we find 
\b
\sigma=\sigma_{\text{geo}} -\frac{4\pi^2}{\omega^2}\text{Re}\left[\sum_{n=1}^\infty \frac{e^{i\pi(\lambda_n -1/2)}\lambda_n\gamma_n |S_{\omega l 0}|^2}{\sin[\pi(\lambda_n-1/2)]}\right],
\label{aga}
\e
where {$\sigma_{\text{geo}}=\pi b_c^2$ ,  and $\gamma_n$ are the residues for the Regge poles $\lambda_n$. For the spheroidal harmonics we consider the approximation used in Ref. \refcite{KerrAbsSca}, given by
\b
|S_{\omega l 0}|^2\approx \frac{1}{2\pi^2}.
\label{osa}
\e

A good approximation for the greybody factors in the high-frequency regime is given by~\cite{Decanini:2011xi}
\b
\Gamma_{\omega lm}\approx \frac{1}{1+e^{-2\pi(\omega-\Omega_0 L)/\Lambda_0}},
\label{gbf}
\e
where $\Omega_0=2\pi/T_0$ is the orbital frequency for the critical orbit and $L\equiv l+1/2$.
By finding the poles of Eq. (\ref{gbf}), which are given by
\b
\lambda_n=\frac{\omega}{\Omega_0} + i\left(n+\frac{1}{2}\right)\beta+\mathcal{O}(1/\omega),
\label{rpl}
\e
we may obtain the corresponding residues, namely
\b
\gamma_n=-\frac{\beta}{2\pi},
\label{res}
\e
where
\b
\beta \equiv \Lambda_0/\Omega_0.
\e

Substituting Eqs. (\ref{osa})--(\ref{res}) in Eq. (\ref{aga}), and using also
\b
\sum_{p=1}^{\infty}e^{i2p\pi(z-a)}=\frac{i e^{i\pi(z-a)}}{2 \sin[\pi(z-a)]}
\e
for $p=1$, we find, from the first Regge pole,~\cite{Decanini:2011xi}
\b
\sigma \approx \sigma_{\text{geo}} \left[1-\frac{8\pi\beta e^{-\pi\beta}}{\Omega_0^2 b_c^2}\text{sinc}\left(\frac{2\pi \omega}{\Omega_0}\right)\right].
\label{sinc}
\e

\section{Numerical results}
\label{sec_nr}

A massless scalar field obeys the Klein-Gordon equation, given by $\square \Psi=0$. To solve this equation we make a separation of variables and write the solution as
\b
\Psi=\sum_{l=0}^{+\infty}\sum_{m=-l}^{+l}\frac{U_{\omega lm}(r)}{\sqrt{r^2+a^2}}S_{\omega lm}(\theta)e^{im\phi-i\omega t}, 
\e
where $S_{\omega lm}(\theta)$ are the oblate spheroidal harmonics figuring in Eqs.~(\ref{abs})--(\ref{osa}). We need to solve the radial equation, given by~\cite{Leite:2017hkm}
\be
\frac{d^2 U_{\omega lm}(r_\star)}{d r_\star^2}&+&\left\{\left(\omega - m\frac{a}{r^2+a^2}\right)^2+
\left[2Mr-2r^2-\Delta+\frac{3r^2}{r^2+a^2}\Delta\right]\frac{\Delta}{\left(r^2+a^2\right)^3}\right.\nonumber\\ 
&-&\left.\left(a^2\omega^2+\lambda_{lm}-2ma\omega\right)\frac{\Delta}{\left(r^2+a^2\right)^2}\right\} U_{\omega lm}(r_\star)=0,
\label{eq:radialeq}
\ee
where $r_\star$ is the tortoise coordinate, defined as
\b
r_{\star}\equiv\int d r\,\left(\frac{r^2+a^2}{\Delta}\right).
\label{eq:tortoisecoord}
\e

In order to solve Eq. (\ref{eq:radialeq}) we need to impose boundary conditions, which we fix as
\be
U_{\omega lm}(r_\star)\sim\left\{
\begin{array}{c l}
	{\mathcal{I}_{\omega lm}}U_I+{{\cal R}_{\omega lm}} U_I^* & (r_\star/M\rightarrow +\infty),\\
	{\mathcal{T}_{\omega lm}} U_T & (r_\star/M\rightarrow -\infty),
\end{array}\right.
\label{inmodes}
\ee  
in which
\be
U_I = e^{-i \omega r_\star}\sum_{j=0}^N \frac{h_j}{r^j},
\label{UI}
\ee
\be
U_T = e^{-i \left({\omega-m\Omega_{\rm H}}\right)r_\star}\sum_{j=0}^N g_j (r-r_+)^j.
\label{UT}
\ee
These are the appropriate boundary conditions to describe a scattering problem, since we have an incoming and a reflected wave at infinity and a transmitted wave at the horizon. $g_j$ and $h_j$ are constants to be determined by substituting the boundary conditions in the radial equation for the corresponding limits. In order to simplify the equations we choose $h_0=g_0=1$. Once we have found the radial solution, we manipulate $U_{\omega lm}$ and $d U_{\omega lm}/dr$ , and find the reflection coefficient, which we use to compute the total absorption cross section, knowing that $\Gamma_{\omega l m}=1-|\mathcal{R}_{\omega l m}/\mathcal{I}_{\omega l m}|^2$.

In Fig. \ref{snf} we compare the results for the sinc approximation, given by Eq. (\ref{sinc}), with numerical results for different choices of $Q$ and $a$. The solid lines are the numerical results, while the dashed lines are the corresponding analytical ones. We see that the sinc approximation presents a good fit to the numerical results, even for relatively small values of the frequency ($\omega M \gtrsim 0.15$).

\begin{figure}
\centering
\includegraphics[width=1\columnwidth]{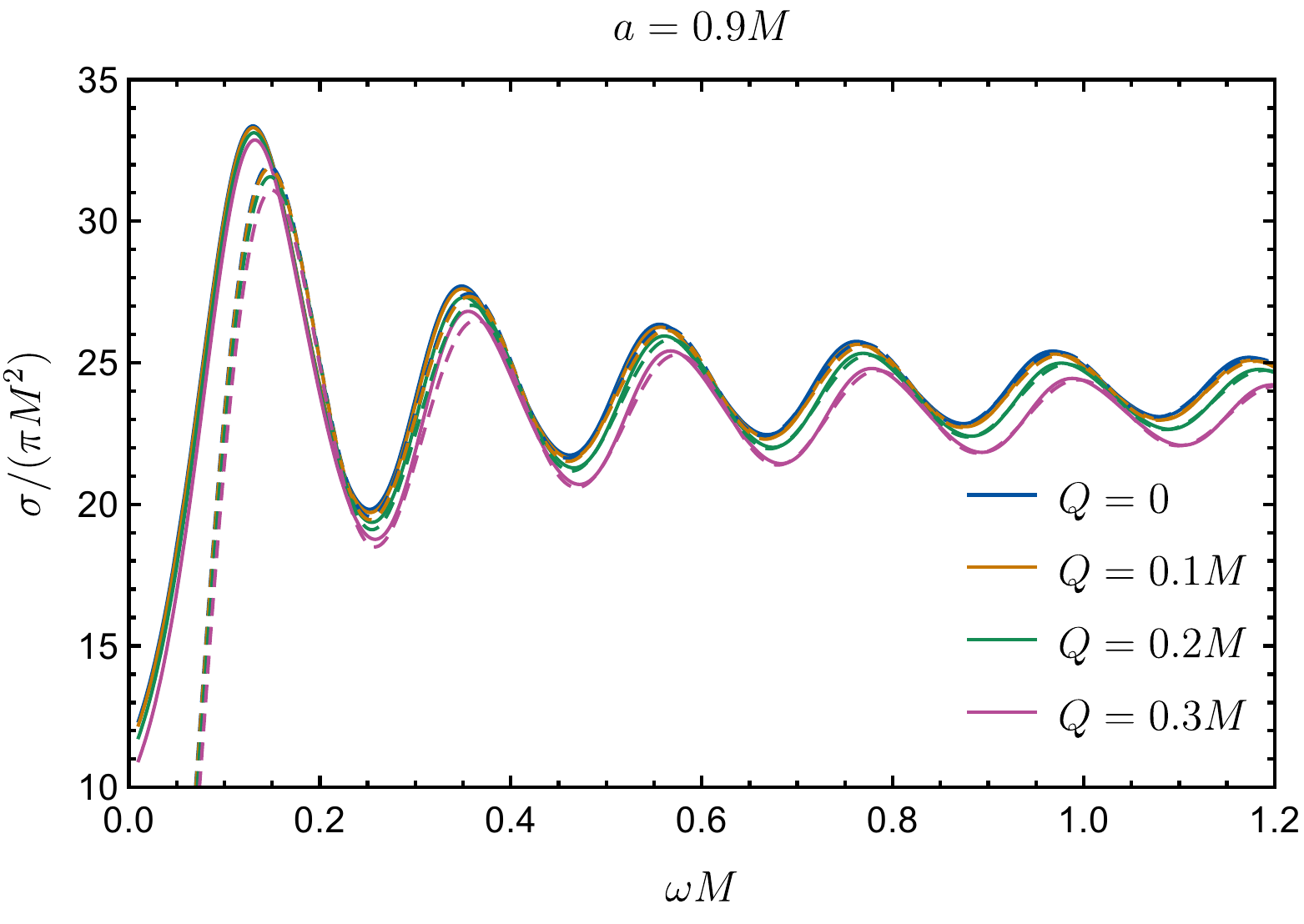}\hspace{0.2in}\\ \vspace{0.3in}
\includegraphics[width=1\columnwidth]{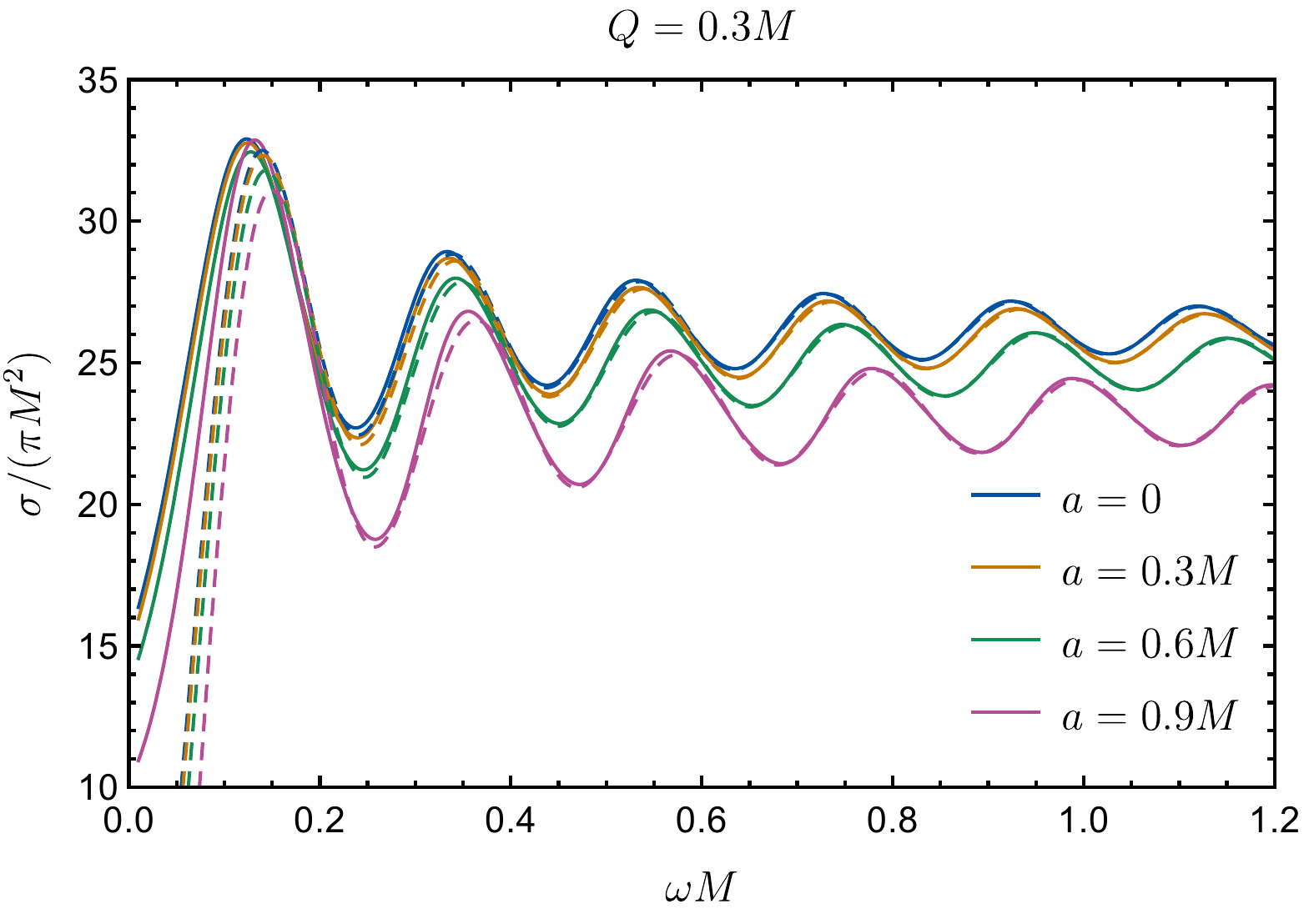}
\caption{Comparison between numerical results and the sinc approximation for the total absorption cross section, considering on-axis impingement. The solid lines are the numerical results, while the dashed lines are the corresponding analytical results obtained through the sinc approximation, given by Eq.~\eqref{sinc}.}
\label{snf}
\end{figure}

\section{Low-frequency approximation}
\label{sec_lf}

The low-frequency limit of the absorption cross section for massless scalar fields is a well known result in the literature, having been found for stationary spacetimes with any dimension \cite{Higuchi:2001si}. In particular, for the Kerr-Newman spacetime it was first obtained by Maldacena and Strominger \cite{Maldacena:1997ih}. 
Here we obtain analytically this low-frequency approximation in a slightly different way. \cite{Unruh:1976, Benone:2016}
We will consider three solutions: 
One for zero frequency ($U^0$), one for $r\rightarrow \infty$ ($U^I$), and one for $r\rightarrow r_+$ ($U^H$). We will then consider the overlapping between these solutions in order to find the reflection coefficient and, hence, the absorption cross section.

We first consider the case for zero frequency. The eigenvalues of the spheroidal harmonics obey the expansion\cite{spheroidal}
\b
\lambda_{lm} = l(l+1)+\sum_{k=1}^\infty c_{2 k }(a\omega)^{2 k}.
\label{clm}
\e
Since we are considering the low-frequency limit, we can take $\lambda_{lm} \approx l(l+1)$. We now solve Eq. (\ref{eq:radialeq}) making $\omega = 0$, for which case we obtain
\b
U^0 = \sqrt{r^2+a^2}A P_l^{\bar{m}}(y)+\sqrt{r^2+a^2}B Q_l^{\bar{m}}(y),
\label{erw}
\e
where $P_l^{\bar{m}}(y)$ and $Q_l^{\bar{m}}(y)$ are the associated Legendre functions of the first and second kind, respectively, $A$ and $B$ are constants and
\b
\bar{m}=-\frac{i a m}{\sqrt{M^2-(a^2+Q^2)}},
\e
\b
y=(r-M)\epsilon,
\e
\b
\epsilon = \frac{1}{\sqrt{M^2-(a^2+Q^2)}}.
\e
In the limit $r\rightarrow r_+$, $y\rightarrow 1$. Since, $Q_l^m(y)\rightarrow \infty$ as $y\rightarrow1$, we then take $B=0$, such that our solution is regular at the horizon.

Now we consider the solution for $r\rightarrow \infty$. In this case Eq.~(\ref{eq:radialeq}) reduces to
\b
\frac{d^2 U_{\omega lm}(r_\star)}{d r_\star^2}+ \left[\omega^2-\frac{l(l+1)}{r^2}\right]U_{\omega lm}(r_\star)=0,
\label{eri}
\e
where we neglected terms of $\omega^2/r^2$ and $\mathcal{O}(1/r^3)$ in Eq.~\eqref{eq:radialeq}. The solution of Eq.~(\ref{eri}) is given by
\b
U^I=\omega r_\star [(-i)^{l+1}\mathcal{I}_{\omega l m}h^*_l(\omega r_\star)+i^{l+1}\mathcal{R}_{\omega l m}h_l(\omega r_\star)].
\label{sri}
\e
The coefficients of Eq.~(\ref{sri}) were chosen such that Eq.~(\ref{sri}) reduces to Eqs.~(\ref{inmodes})-(\ref{UT}) with only the $j=0$ term retained. Indeed, in the limit $\omega r_\star \gg l(l+1)$, we have
\b
h_l(\omega r_\star)=(-i)^{l+1}\frac{e^{i\omega r_\star}}{\omega r_\star}.
\label{hil}
\e
Substituting Eq. (\ref{hil}) in Eq. (\ref{sri}), we obtain
\b
U^I=\mathcal{I}_{\omega l m}e^{-i\omega r_\star}+\mathcal{R}_{\omega l m}e^{i\omega r_\star}.
\e

We will now consider the overlapping between Eqs. (\ref{erw}) and (\ref{sri}). In order to do that, we first consider the limit $r\rightarrow \infty$, for which case the associated Legendre functions reduce to
\b
P_l^m(x)=\frac{(2l)!}{2^l i^m l! (l-m)!}x^l,
\e
such that Eq. (\ref{erw}) reduces to
\b
U^0=A\frac{(2l)!}{2^l i^{\bar{m}} l! (l-{\bar{m}})!}\epsilon^l r^{l+1},
\label{ewi}
\e
where $(l-\bar{m})!=\Gamma(l-\bar{m}+1)$.

Considering now the limit of Eq. (\ref{sri}) for $\omega r_\star \ll 1$, we obtain
\be
U^I &=& [(-i)^{l+1}\mathcal{I}_{\omega l m}+i^{l+1}\mathcal{R}_{\omega l m}]\frac{2^l l!}{(2l+1)!}(\omega r_\star)^{l+1}\nonumber\\
&+& i[(-i)^{l+1}\mathcal{I}_{\omega l m} - i^{l+1}\mathcal{R}_{\omega l m}]\frac{(2l)!}{2^l l!}(\omega r_\star)^{-l},
\label{eiw}
\ee
where we used that $h_l(x) = j_l(x)+i n_l(x)$, with
\b
j_l(x) = \frac{2^l l!}{(2l+1)!}x^l,
\e
\b
n_l(x) = -\frac{(2l)!}{2^l l!}x^{-(l+1)}.
\e
In order to guarantee that the solution is not divergent in the low-frequency limit, we assume $\mathcal{R}_{\omega l m} \approx (-1)^{l+1}\mathcal{I}_{\omega l m}$, such that Eq. (\ref{eiw}) gives us
\b
U^I=(-i)^{l+1}\mathcal{I}_{\omega l m}\frac{2^{l+1} l!}{(2l+1)!}(\omega r_\star)^{l+1}.
\label{eiw2}
\e

Comparing Eq. (\ref{ewi}) and Eq. (\ref{eiw2}), we find that
\b
A=\frac{(-i)^{l+1}2^{l+1+\bar{m}} (l!)^2(l-\bar{m})! \omega^{l+1}}{(2l)!(2l+1)!\epsilon^l}\mathcal{I}_{\omega l m},
\e
such that Eq. (\ref{erw}) is given by
\b
U^0 = \frac{(-i)^{l+1}2^{l+1+\bar{m}} (l!)^2(l-\bar{m})! \omega^{l+1}}{(2l)!(2l+1)!\epsilon^l}\mathcal{I}_{\omega l m}\sqrt{r^2+a^2} P_l^{\bar{m}}(-M\epsilon +\epsilon r).
\label{erw2}
\e

Let us now consider the case for $r\rightarrow r_+$. Taking $j=0$ in Eq.~(\ref{UT}), we obtain
\b
U^H = \mathcal{T}_{\omega l m} e^{-i(\omega - m\Omega_H)r_\star}.
\label{erh}
\e
Considering the case for $\omega \ll 1$, we find
\b
U^H = \mathcal{T}_{\omega l m}.
\label{ehw}
\e

We now consider the limit $r\rightarrow r_+$ in Eq.~(\ref{erw2}). 
In order to simplify our calculations we consider only the case $m=0$, for which we have $P_l^0(1)=1$. Substituting this result in Eq.~(\ref{erw2}), we have
\b
U^0 = \frac{(-i)^{l+1}2^{l+1} (l!)^2l! \omega^{l+1}}{(2l)!(2l+1)!\epsilon^l}\mathcal{I}_{\omega l 0}\sqrt{r_+^2+a^2}.
\label{ewh}
\e
Comparing Eqs. (\ref{ehw}) and (\ref{ewh}), we find
\b
\frac{\mathcal{T}_{\omega l 0}}{\mathcal{I}_{\omega l 0}} = \frac{(-i)^{l+1}2^{l+1} (l!)^2l! \omega^{l+1}}{(2l)!(2l+1)!\epsilon^l}.
\e
Taking $l=m=0$, we find
\b
\frac{\mathcal{T}_{\omega 0 0}}{\mathcal{I}_{\omega 0 0}} = -i 2 \omega \sqrt{r_+^2+a^2}.
\label{tlf}
\e
Now we have to substitute Eq.~(\ref{tlf}) in Eq.~(\ref{abs}). However, as it can be also inferred from Eq.~(\ref{clm}), the difference between the spheroidal harmonics and the spherical harmonics comes from terms which depend on $a\omega$. Thus, in the limit $\omega \rightarrow 0$, $S_{\omega lm}(\theta)\rightarrow Y_l^m(\theta)$, with $Y_l^m$ being the scalar spherical harmonics, which can be written as 
\b
Y_l^m(\theta)=\sqrt{\frac{(2l+1)}{4\pi}\frac{(l-m)!}{(l+m)!}}P_l^m(\cos\theta)\label{eq:spherical_harmonics}.
\e
Substituting Eq.~(\ref{tlf}) and Eq.~\eqref{eq:spherical_harmonics} with $l=m=0$ in Eq.~(\ref{abs}), we obtain
\b
\sigma =4 \pi (r_+^2+a^2),
\e
which is the area of the black hole event horizon.

\section{Final remarks}
\label{sec_fr}

We considered null geodesics with incidence along the rotation axis of a charged rotating black hole. We solved the equations for the critical geodesic, finding analytical expressions for the geodesic parameters, such as the Lyapunov coefficient. We used the Regge pole technique to find a closed form for the high-frequency limit of the total scalar absorption cross section for the on-axis case, the so-called {\it sinc approximation}. 
We presented a selection of numerical results, obtaining excellent agreement with the corresponding analytical 
approximations, even in the case of relatively small frequencies.
We also obtained an analytical approximation for the total absorption cross section in the  low-frequency limit. 

\section*{Acknowledgments}
The authors would like to acknowledge 
Conselho Nacional de Desenvolvimento Cient\'ifico e Tecnol\'ogico (CNPq)
 and Coordena\c{c}\~ao de Aperfei\c{c}oamento de Pessoal de N\'ivel Superior (CAPES) 
 for partial financial support.
S.D.~acknowledges financial support from the Engineering and Physical Sciences Research Council (EPSRC) under Grant No.~EP/M025802/1 and from the Science and Technology Facilities Council (STFC) under Grant No.~ST/L000520/1.


\end{document}